\documentclass[preprint]{aastex}
\usepackage{emulateapj5}
\input epsf

\def\ion#1#2{#1\,{\sc #2}}

\newcommand{\deltE}{\Delta\kern-1ptE}
\newcommand{\fuse}{{\it FUSE}}
\newcommand{\kms}{km~s$^{-1}$}
\newcommand{\gla}{$\lambda$}

\begin{document}
\submitted{1 June 2000}
\shorttitle{\fuse\ OBSERVATIONS OF AB DORADUS}
\shortauthors{AKE ET AL.}
\title{\fuse\ Observations of the Active Cool Star AB Doradus}

\author{T. B. Ake}
\affil{Department of Physics \& Astronomy, The Johns Hopkins University, and 
Computer 
Sciences Corporation, 3400 N. Charles Street, Baltimore MD, 21218}

\author{A. K. Dupree and P. R. Young}
\affil{Harvard-Smithsonian Center for Astrophysics, Cambridge MA
02138}


\author{J. L. Linsky}
\affil{JILA, University of Colorado and NIST, Boulder CO 80309-0440}

\author{R. F. Malina, N. W. Griffiths, and O. H. W. Siegmund}
\affil{Space Sciences Laboratory, University of California, Berkeley, Berkeley 
CA 94720} 

\and

\author{B. E. Woodgate}
\affil{Laboratory for Astronomy and Solar Physics, NASA's GSFC, Code 681, 
Greenbelt MD 20771}

\begin{abstract}
Far ultraviolet spectra were obtained of the 
active cool star AB Doradus (HD 36705) during 
the calibration and checkout period of the \fuse\ satellite. Observations in
this early  phase of the mission were taken at a resolving power of 12000--15000
($\sim$20--25 \kms) and  covered the spectral range 905--1187\AA. The
integrated spectrum exhibits  strong, rotationally broadened stellar
emission from \ion{C}{iii} (\gla977, \gla1175) and \ion{O}{vi} (\gla1032,
\gla1037), and 
many weaker lines.  Strong emission lines of \ion{C}{iii}
and \ion{O}{vi} exhibit broad wings. 
The \ion{C}{iii} $\lambda$977 profile
shows blue-shifted absorption at $\sim$30 \kms and \ion{C}{ii} \gla1036 
absorption appears superposed on emission in the wing of \ion{O}{vi} 
\gla1037. 
Rotational modulation of \ion{C}{iii} and \ion{O}{vi} is
present, in harmony with its photometric variability.
Flares were detected in the brightest
lines and subexposures  were analyzed to examine flux and profile
variations. Downflows that extend to 600 \kms\  during a flare are found
in the \ion{O}{vi} profiles.   These early  observations
demonstrate that \fuse\ will be an exceptional instrument for
studying chromospheres in cool stars.
\end{abstract}

\section{Observations and Data Analysis}

\begin{figure*}[ht!]
\vspace{-0.3in}
\epsfxsize=6.5in
\epsfbox{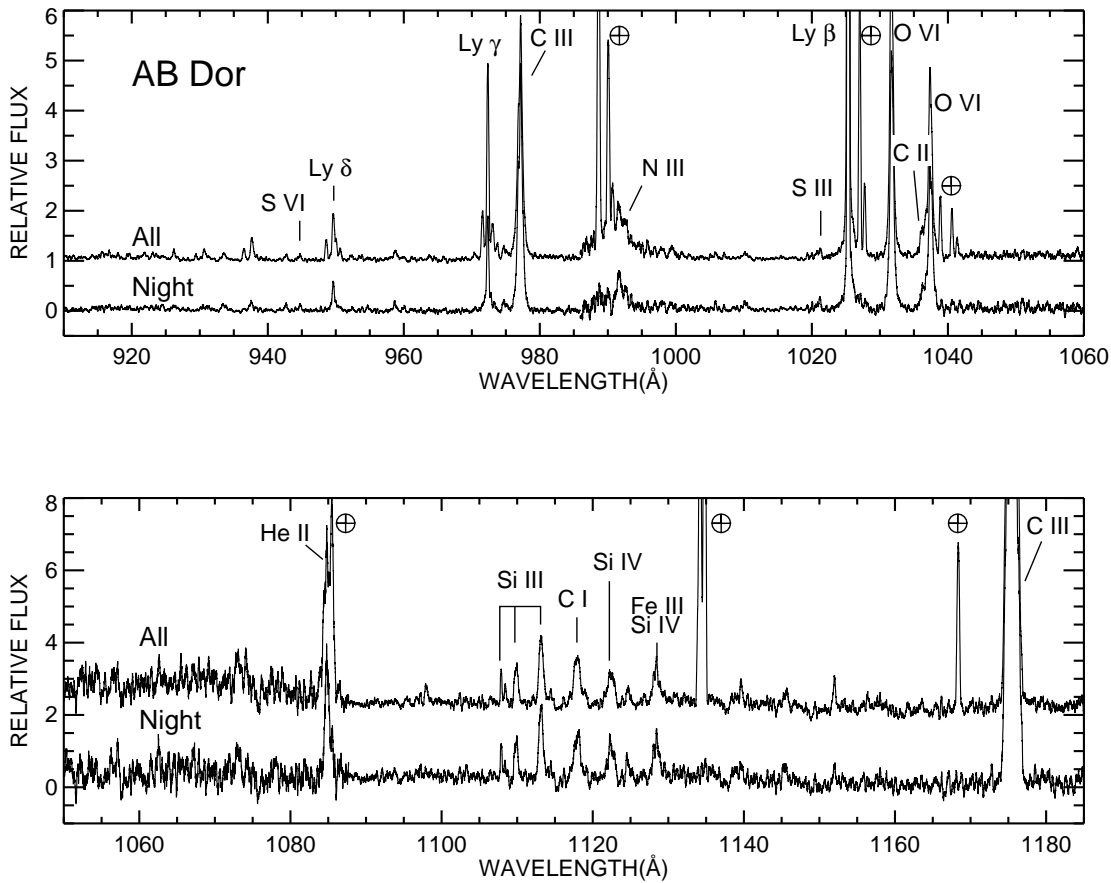}
\figcaption{\fuse\  spectra (SiC1A, SiC1B, SiC2B, LiF1B, LiF2B) 
of AB Dor taken  on 1999 December 14  with prominent stellar
and airglow emission features identified. The upper spectrum in each panel 
includes photons collected during orbital day and night; the lower spectrum is 
derived from night time photons only. Rotational broadening is 
apparent even in  these compressed plots.
}
\end{figure*}

AB Doradus is a nearby (15 pc), rapidly rotating (P$_{rot}$ = 12.4 hr, 
$v$~sin~$i$ = 90
\kms),  K0--2 IV--V pre-main sequence star that exhibits high activity
and frequent  flaring (e.g., Collier Cameron et al. 1999;
Schmitt, Cutispoto \& Krautter 1998;
Vilhu et al. 1998). It is  the first cool star observed by
the {\em Far Ultraviolet Spectroscopic Explorer} (\fuse, cf. Moos et al.
2000). Observations were obtained on 1999  October 20 and 22,
in the LiF channels (\gla\gla987--1187), and on 1999  December 14, when
exposures in all four channels were
made covering the range \gla\gla905--1187.  Exposures were taken in
time-tagged mode at 1 second resolution through the 30$''\times$30$''$
aperture and were scheduled when AB Dor was in the 
continuous viewing zone. Thus good coverage was obtained over most of its 
rotational period each day. The only interruptions were due to South Atlantic 
Anomaly crossings, when detector high voltage is reduced.
The total exposure time amounted to 79 ks for the LiF channels and 24 ks for 
SiC, of which 33 ks and 13 ks respectively were 
obtained during orbital night. The spectral resolving power was 12000--15000 
($\sim$20--25 \kms) but the considerable oversampling improves the spectral 
purity.

Since the spectrographs were not fully focused, care was exercised when 
comparing lines from different channels.   
Data sets were reprocessed with the calibration pipeline as improvements 
became available.  Both airglow and the scattered solar spectrum can 
contribute emission lines expected to be present in cool stars; thus spectra 
obtained 
from data taken during orbital night  were examined and analyzed 
separately in regions where the contamination was significant. For time 
variability
studies, subexposures were created from the photon lists and reprocessed 
individually. 

Wavelength calibration errors especially near the  
ends of detector segments are still present. For AB Dor, the
easily identifiable  lines and airglow lines allowed for local
adjustments of the scale, but  identifications for weaker lines had to
be confirmed through comparisons with  solar disk spectra from SUMER
(e.g., Curdt et al. 1997). 
The wavelength scale  uncertainty 
precludes at present an analysis of possible differential motions between  the
lines.

\section{Line Spectrum}

The FUSE spectrum of AB
Dor is shown in Figure 1. The strongest 
lines  arise from \ion{C}{iii} (\gla977, \gla1175), \ion{O}{vi}
(\gla1032, \gla1037) and  airglow 
lines of \ion{H}{i}, \ion{O}{i}, \ion{N}{i}, and \ion{N}{ii}. 
Most of the latter can be distinguished from  stellar components by
screening out parts of the exposures taken in daylight as shown
in Fig. 1. We can
tentatively identify fainter lines due to \ion{H}{i}, \ion{He}{ii}, \ion{C}{i},
\ion{C}{ii}, \ion{N}{iii}, \ion{Si}{iii}, \ion{Si}{iv}, \ion{S}{iii}, 
\ion{S}{vi}, and \ion{Fe}{iii}, which will be discussed in a 
later paper.
The lines are broadened by the high rotation rate of the
star.

Profiles of chromospheric and transition region lines can 
provide clues to heating mechanisms and mass motions in
cool stars. With a resolving power of 12000--15000, \fuse\ can be used to study
thermal and non-thermal motions, which  typically range from 15 to
hundreds of \kms\ in these stars (e.g., Wood, Linsky, \& Ayres 1997).  
Figure 2 shows the \ion{O}{vi} doublet region in detail for the time-integrated 
LiF1A spectrum on October 22. The 
$\lambda$1032 component is cleanly separated from other features in
the spectrum  and is symmetrical.  A Gaussian profile broadened by
90 \kms\ rotation with no limb darkening 
can adequately reproduce the core of the line, but does not fit the wings.
\citet{schmitt97} were able to measure the rotational broadening in \ion{O}{vi} 
with {\it ORFEUS}, but could not adequately discern the broad wings.
 The 
low scattered
light properties of \fuse\ \citep{sahnow00} and the relatively modest FUV
count rates for AB Dor make an instrumental source
unlikely to explain the extended profile. Extra
flux in the wings of \ion{Si}{iv} and \ion{C}{iv} lines has been observed in 
other active stars \citep{wood97}, and the \ion{O}{vi} observations of AB Dor 
extend this phenomenon to greater line-formation temperatures. 
Since telescope focus and spectrograph calibration had not been completed, there 
is some disagreement in the details of the line profiles between the \fuse\ 
channels. Considering all the data, 
we find a two-Gaussian
fit composed of a narrow (FWHM $\simeq 80$ \kms)
and broad (FWHM $\simeq 300$ \kms) component, the latter contributing 38--56\% 
of the
total line flux depending on the data set, reproduces the
\gla1032 profile when rotationally and instrumentally 
broadened. These values are comparable to
widths of 68 and 334 \kms\ determined for \ion{C}{iv} in AB Dor from
GHRS spectra \citep{vilhu98}. \citet{wood97} found
broad components in a number of cool stars to be 
correlated with \ion{C}{iv} and X-ray surface fluxes, and 
proposed that their origin is microflare heating.  However
extended material may occur frequently in rapidly rotating
cool stars.  Co-rotating
prominence material or an extended chromosphere as \citet{cc89} 
have identified on AB Dor might cause similar broadening. 
Another rapidly rotating K dwarf, V471 Tau, exhibits
material at 10$^5$K extending to 1 $R_\star$ above the surface 
\citep{guinan86}.

The weaker member of the \ion{O}{vi} doublet ($\lambda$1037.61) is blended
with \ion{C}{ii} emission (\gla1036.34, \gla1037.02) in its blue wing. 
A strong \ion{C}{ii} \gla1036 absorption feature is superposed on the emission,
likely due to interstellar or circumstellar material.  
Except for the blue wing, 
the \ion{O}{vi} $\lambda$1037 line is well matched by the same Gaussian fit from
$\lambda$1032 when shifted by the difference in wavelengths and scaled by
the doublet ratio.

\ion{C}{iii} $\lambda$977 differs from \ion{O}{vi} by additional
absorption appearing $\sim$30 \kms\ from line center 
and irregularities in the wings of the line profile (Figure 3). We have 
considered several possible origins for this, which cannot be resolved until a 
better wavelength scale is available.   
A similar asymmetric profile in \ion{C}{iii} was
found in the luminous cool star $\beta$ Dra from {\it ORFEUS} spectra
where it was attributed to a wind \citep{dupree98}.
AB Dor frequently displays transient absorption features in the H$\alpha$
profile \citep{cc99}, and the \ion{C}{iii} absorption
may be a high temperature manifestation of the behavior
in H$\alpha$. Finally, the presence of the \ion{C}{ii} \gla1036 absorption
line suggests that the \ion{C}{iii} absorption could also be  
interstellar or circumstellar in origin.

\section{Electron Densities}
We compute a ratio of the density sensitive \ion{C}{iii} lines of,
\gla1175/\gla977 = 0.651 $\pm$0.172 (in erg units), although 
the absolute and relative 
flux calibration of \fuse\ is still at an early stage.  This
ratio, when compared to the  calculations of the
CHIANTI code, suggest that the density at T$\sim$80000 K is near the
high density limit, namely $\sim10^{11}$ cm$^{-3}$, and
similar to the values derived from {\it ORFEUS} spectra of AB Dor 
\citep{schmitt98}. However the uncertain calibration
and presence of absorption in the
\gla977 line, potentially weakening the
line and increasing the inferred density, 
suggests that caution must be used in applying this diagnostic.

\section{UV Flares on AB Dor}

AB Dor undergoes frequent flaring, particularly in X rays; impulsive
flares have  also been detected in \ion{C}{iv} \citep{vilhu98}. Several flares 
occurred during the \fuse\
observations; one was bright enough on 1999 October 22  to extract a
spectrum in the strongest lines. Figure 4 shows a
dynamic spectrum in the \ion{C}{iii} $\lambda$1175 region.
Light curves for \ion{C}{iii} \gla1175 and \ion{O}{vi} show
similar temporal behavior marking a short-duration 
($\approx$700 sec)
flaring episode.   The overall brightening of 
the $\lambda$1175  line was accompanied 
by an enhanced red
shifted component.  Here care was taken in determining which
features are stellar flares as distinguished from events bursts seen with
the detector \citep{sahnow00}. Fig. 2  shows the spectrum of  
red shifted components in the \ion{O}{vi} lines;
the Doppler  shift of a Gaussian fit to the flare component  is
centered at +235 \kms, with extension to $+$600 \kms.  A
flare in \ion{C}{iv} observed by \citet{vilhu98} exhibited a broad blue-shifted
Gaussian profile superposed on the quiescent profile and was attributed to 
evaporation of chromospheric
material.   Flares on
AD Leo have shown high redshifts 
in \ion{C}{iv}, amounting to $\sim$650 \kms\ 
with extensions to 1800 \kms\ \citep{bb92}. Downward
flowing flare 
material is a frequent occurrence in solar post-flare magnetic 
loops and could explain the profiles observed here.

\section{Rotational Modulation}

AB Dor undergoes optical and UV
variations as  active structures rotate through the line of
sight.  Maximum spottedness occurs  at minimum light in V and U. In
Figure 5, we show the variation of the strongest 
lines (\ion{C}{iii} and \ion{O}{vi}) on 
1999 October 22, when AB Dor appears to have been most active, 
by dividing the \fuse\ observations  into subexposures 
of $\sim$35 min apiece. 
Rotational phase is determined from the ephemeris of \citet{innis88}:
HJD =  2444296.575 + 0.51479~E. In this case,
minimum light occurs at  $\phi$ = 0.5. Photometry in 1999 December
from Mt.\ Kent Observatory in Queensland 
confirms  that AB Dor continues to be faint at $\phi$=0.5 on this
ephemeris, although coverage over the entire rotational period was not obtained.

A peak in the \fuse\ fluxes occurs near $\phi\sim$0.6; this 
would  be expected from enhanced high temperature 
activity due to star spots on the photosphere and
overlying regions of activity.  Comparable modulation
amplitudes in \ion{C}{iii} and \ion{O}{vi} suggest the spatial distribution
of the corona is similar between 80000 K and 2$\times$10$^5$~ K
where these lines are formed.   Greater enhancement of
\ion{C}{iii} than \ion{O}{vi} during the flare may be affected by
the increased density enhancing the metastable level of \ion{C}{iii}.  
The large flare on this
date,  however, occurs after the photometric
minimum, near $\phi$ = 0.9 suggesting
that  it arose from
an active region  not associated with the maximal spot coverage.

Simultaneous {\it ORFEUS} and {\it ROSAT} observations are discussed by 
\citet{schmitt98}, who found that FUV and X-ray variations are correlated but do 
not appear to be rotationally modulated, except at about the 15\% level for 
X-ray emission.  AB Dor is known to undergo night-to-night optical variations 
\citep{cc99}, so \ion{O}{vi} rotational modulation could be a transient 
phenomenon dependent on the activity of the star. For the X-ray emission, either 
the coronal material is more homogeneously distributed than the lower 
temperature material or the activity is at higher latitudes, and hence, always 
visible. We note that \citet{cc99} find AB Dor has a stable polar cap of 
photospheric spots.

\section{Conclusions}

This first set of observations of a cool star demonstrates that
\fuse\ can provide insight  into a variety of phenomena 
on AB Dor. We have identified lines
spanning more than a decade  in line formation temperature that can be used to
determine diagnostics in the far ultraviolet. Even though the
instrument is not fully focused, the line  profiles can be studied in
detail, and we have begun examining the dynamics and  heating
processes in this star. Through time-resolved spectroscopy, we can
study  variability on different time scales, from seconds to
days. Importantly, the  redundant wavelength coverage from different
channels provides high confidence  in observed variations of features and
structures.

\acknowledgements
This work is based on data obtained for the Guaranteed
Time Team by the NASA-CNES-CSA FUSE mission operated 
by The Johns Hopkins University.
Financial support to U.S. participants has been provided by
NASA Contract NAS5-32985.
Photometric observations during 1999 December were 
kindly provided by Ian Waite and Brad Carter of 
the University of South Queensland, Australia.

\clearpage

\begin{figure*}[ht!]
\centerline{\epsfxsize=5.in\epsfbox{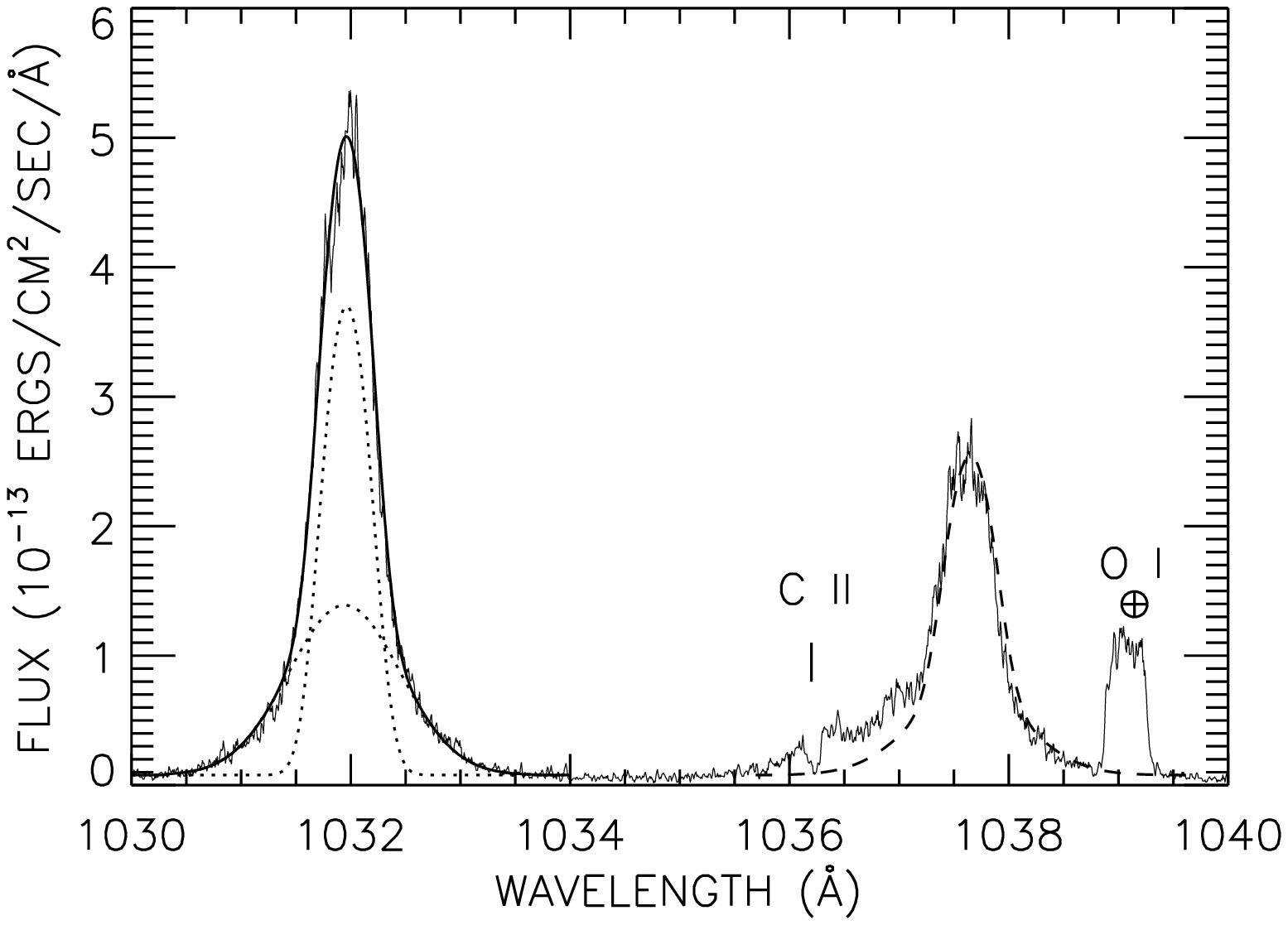}}
\vspace{+0.2in}
\centerline{\epsfxsize=5in\epsfbox{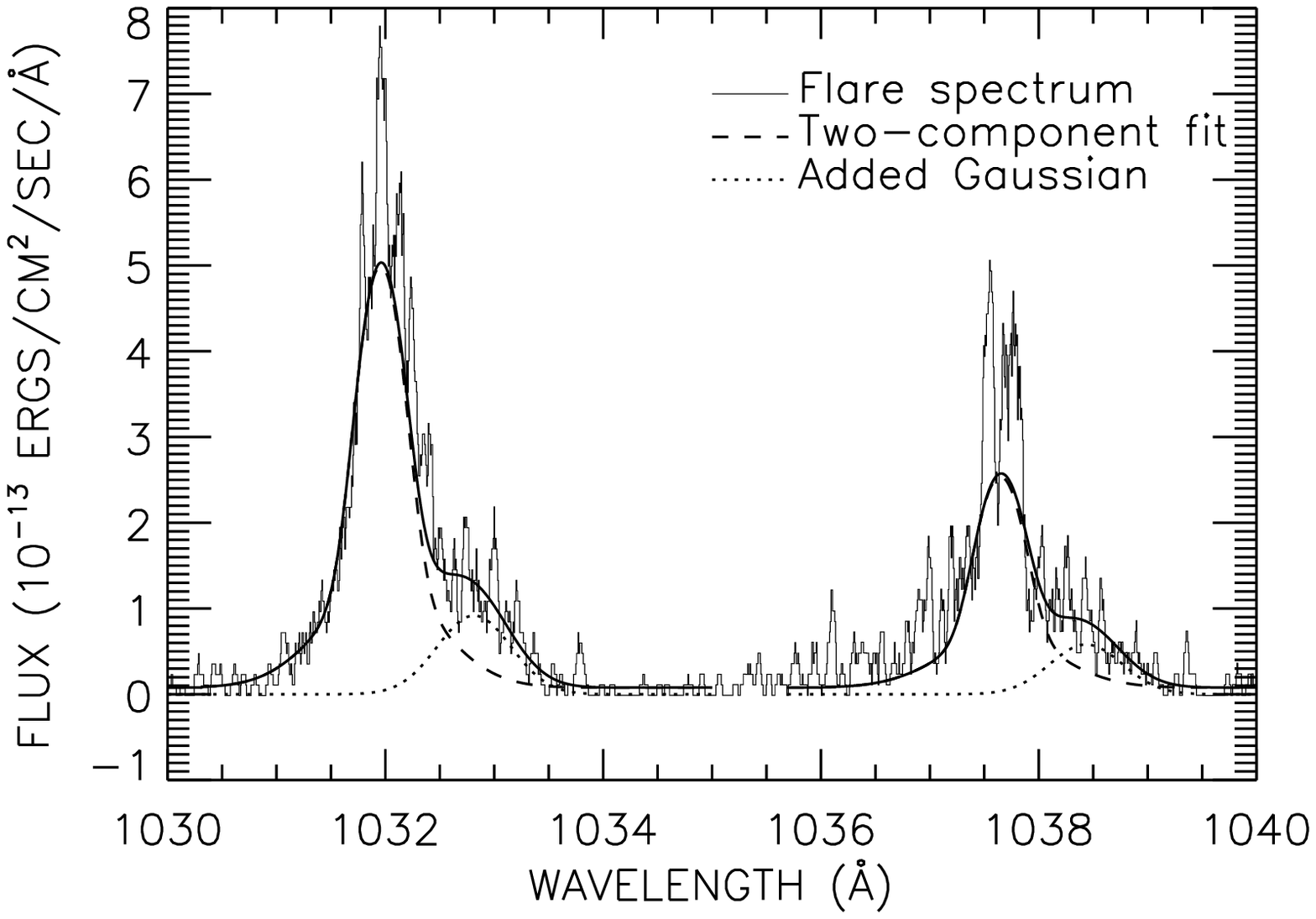}}
\vspace{0.2in}
\caption{{\it Upper panel:} The \ion{O}{vi} 
doublet, time-integrated over the 1999 
October 22 exposures with the LiF1A channel. A two-component Gaussian fit is 
required for the
$\lambda$1032 line. Shown are the Gaussians after rotational and
instrumental broadening (dotted  lines), the 
final fit (thick line), and the same
profile shifted and scaled by a  factor of two for $\lambda$1037
(dashed line). Stellar \ion{C}{ii} emission is noted shortward
of \ion{O}{vi} \gla1037, with interstellar or circumstellar
\ion{C}{ii} $\lambda$1036 absorption  superimposed. 
{\it Lower panel:} Flare profiles are shown averaged over a
2500 s active period of the star. Red shifted material at +235 \kms, extending 
to 600 \kms, 
is found in \ion{O}{vi}, 
as well as sharp emission components near line
center. Dashed  line is the two-component fit from the  time-integrated
profile in the upper panel.} 
\end{figure*}
\clearpage

\begin{figure*}
\vspace{-2.in}
\centerline{\epsfxsize=5.in\epsfbox{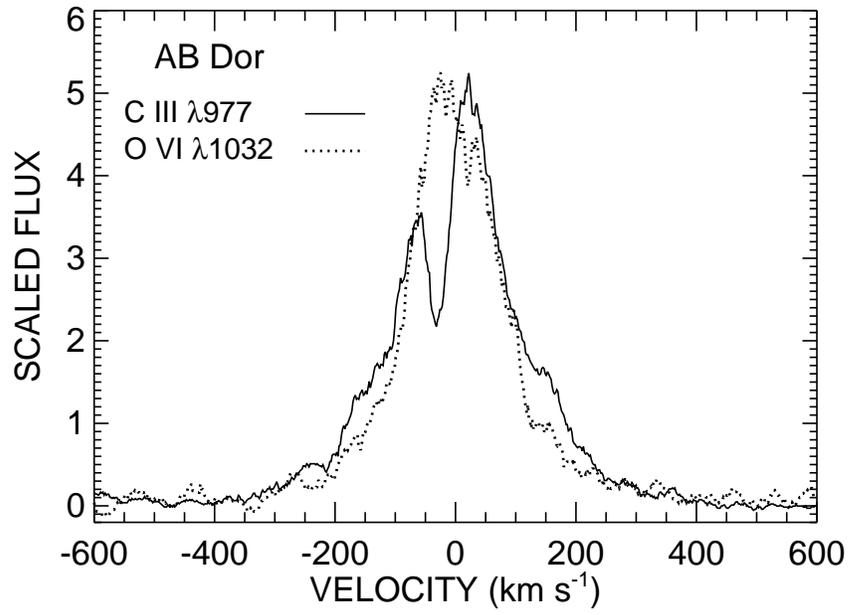}}
\figcaption{Comparison of profiles of the \ion{O}{vi} and
\ion{C}{iii} lines
showing the short wavelength absorption feature in \ion{C}{iii},
and broadened ``bumpy'' wings of emission in the \ion{C}{iii} 
profile.  The velocity scale has an arbitrary offset for each
line, and was simply coaligned in this figure.} 
\end{figure*}
\clearpage

\begin{figure*}
\centerline{\epsfxsize=5.in\epsfbox{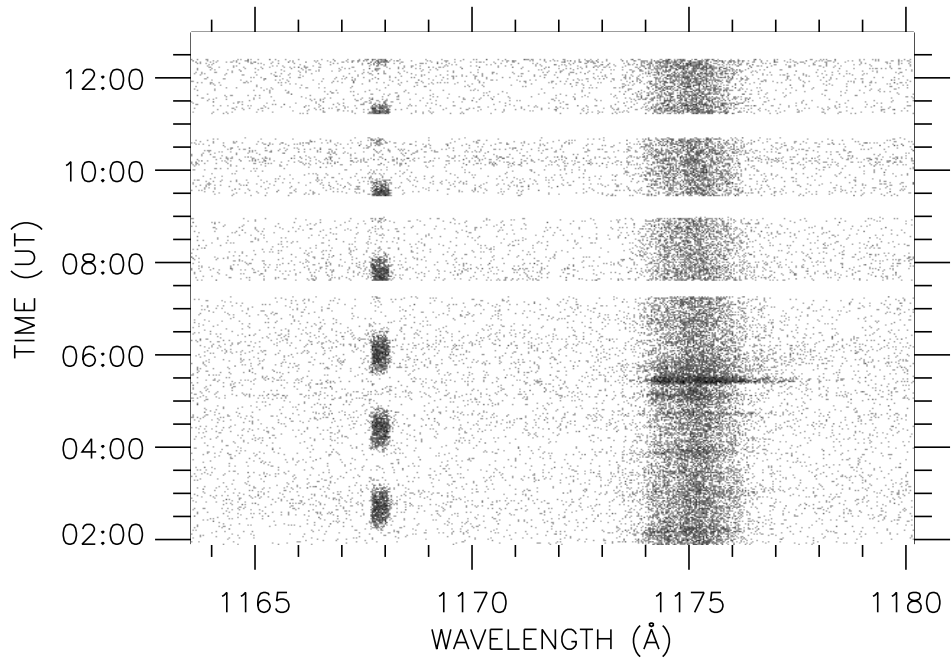}}
\figcaption{Dynamic spectra of \ion{C}{iii} $\lambda$1175 
on 1999  October 22 totaled in 10 
second bins. Small precursor flares occurred before the 
bright event at 05:28UT where both increased flux and
red shifted emission are found. 
Gaps in data are due to SAA passages. The feature showing 
diurnal variation at 
$\lambda$1168 is second-order scattered solar He I.}
\end{figure*}

\begin{figure*}
\centerline{\epsfxsize=5.in\epsfbox{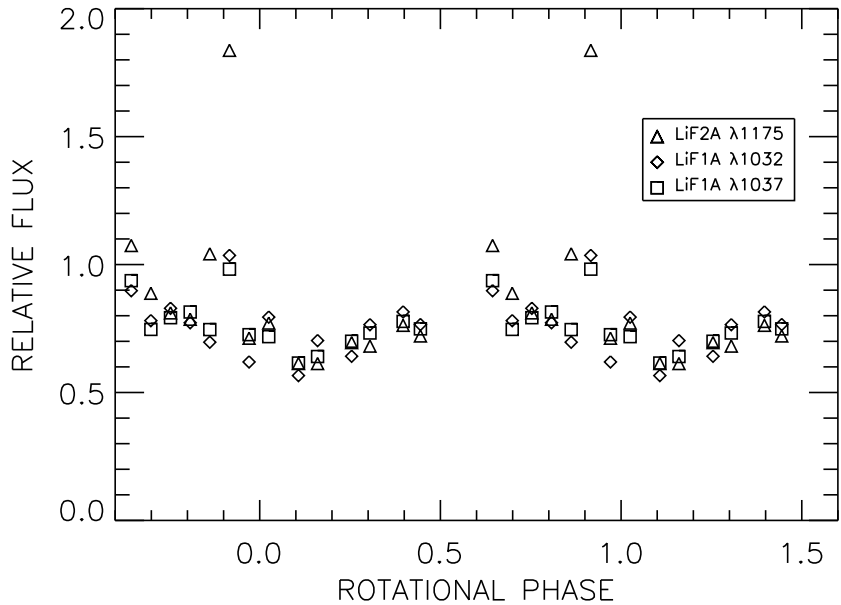}}
\figcaption{Variation of the \ion{O}{vi} \gla1032, \gla1037 
and \ion{C}{iii} \gla1175 
line fluxes with rotational phase for 1999 
October 22 observations binned in 35 minute intervals. The
measurements are replicated over two rotational phases.
Spot maximum is expected at
$\phi\sim0.5$. The large flare  at  05:28 UT occurred at $\phi=0.9$.}

\end{figure*}

\end{document}